\documentclass[aps,prl,letterpaper,twocolumn,nofootinbib,floatfix]{revtex4}

\usepackage{amsfonts}
\usepackage{amsmath}
\usepackage{amssymb}
\usepackage{graphicx}
\usepackage{comment}
\usepackage{times}
\usepackage{setspace}

\begin{document}

\title{Comment on "Characterization of the tunneling conductance across DNA bases"}

\renewcommand{\thefootnote}{\fnsymbol{footnote}}
\author{$^{1}$Johan Lagerqvist, $^{2}$Michael Zwolak, and $^{1}$Massimiliano Di
Ventra\footnote{E-mail address: diventra@physics.ucsd.edu}}
\affiliation{$^{1}$Department of Physics, University of California,
San Diego, La Jolla, CA 92093-0319} \affiliation{$^{2}$Physics
Department, California Institute of Technology, Pasadena, CA 91125}
\date{\today}
\begin{abstract}
In a recent article,  Zikic {\it et al.} [Phys. Rev. E {\bf 74},
011919 (2006)] present first-principles calculations of the DNA
nucleotides' electrical conductance. They report qualitative and
quantitative differences with previous work, in particular with that
of Zwolak and Di Ventra [Nano Lett. {\bf 5}, 421
(2005)] and Lagerqvist {\it et al.} [Nano Lett. {\bf 6}, 779
(2006)]. In this comment we address the alleged discrepancies, showing that
they come from a misrepresentation of our research. Further, we discuss in more
detail the
issue of geometric fluctuations previously investigated by us and raised again
in the work of Zikic {\it et al.} In addition, we point out erroneous comments
made by Zikic {\it et al.} regarding the use of DFT calculations in transport.
\end{abstract}

\maketitle

Recently, Zikic {\it et al.}~\cite{Zikic2006-1} report the
conductance of passivated DNA nucleotides located in between
nanoscale gold electrodes using density-functional theory (DFT)
within the known exchange-correlation (xc) functionals. In several
places throughout their article, they compare their findings with
previously published results by
us~\cite{Zwolak2005-1,Lagerqvist2006-1} and conclude that there are
both qualitative and quantitative differences with their work. We
point out that Zikic {\it et al.} misrepresent the existing
literature by leaving out important details and, further, make
comparisons which are at odds with their own approach and
conclusions. Also, some comments in their work raise general
questions about the adequacy of static approaches to transport and
the differences between such approaches. We address these issues
below.

Zikic {\it et al.} correctly state that the electronic signature of
nucleotides is strongly dependent on what they call ``geometrical
factors''. Their work is an explicit demonstration of a well known
concept: the tunneling current depends exponentially on the width of
the tunneling barrier, which is here formed by the reduced coupling
between the electrodes and the nucleotide. For the case of DNA
between electrodes, this means that changes in nucleotide
orientation modifies their coupling to the electrodes, and therefore
can drastically change the electrical conductance. In addition, if
one fixes the sugar-phosphate backbone position, the different sizes
and geometries
of the bases will cause them to be more or less close to the
electrodes and therefore cause a difference in their relative
conductance. Zikic {\it et al.} seem to indicate that these
conclusions are qualitatively and quantitatively different from
ours. Instead, we understood this fact and we stated explicitly in
Ref.~\onlinecite{Zwolak2005-1} that ``[how well the HOMO and LUMO]
states couple to both electrodes determines the overall magnitude of
the relative currents [between the bases]''. In addition, well prior
to their work, recognizing the importance of geometrical factors, we
explored this issue in much more detail by investigating realistic
structural fluctuations in Ref.~\cite{Lagerqvist2006-1}.

Due to the importance of geometry in the conductance (and relative
conductance) of the different nucleotides, one can not
quantitatively compare the results of Ref.~\onlinecite{Zikic2006-1}
with those of Refs.~\onlinecite{Zwolak2005-1,Lagerqvist2006-1}. In
addition, Zikic {\it et al.} state that their DFT calculations give
what they call mutually consistent results with different
exchange-correlation functionals. By {\it mutually consistent} they
mean that the ordering of the current magnitudes are the same
regardless of the xc-functional used. Consistency may hold true for
the current averaged over all their configurations and at small
bias, but it is obvious from their own work that it does not hold
true otherwise. For instance, by examining either the conductance or
the current in their Figures 8, 9, 10, 11, or 12, one can extract
essentially any desired ordering in the nucleotides' conductance.

Thus, we point out that it is incorrect for them to claim that our
results in Ref.~\onlinecite{Zwolak2005-1} are not consistent with
our results in Ref.~\onlinecite{Lagerqvist2006-1} based on the
change in conductance ordering. This claim leaves out crucial facts
which are clearly written in our papers: from one paper to the other
we did change {\em i)} the bias, {\em ii)} the electrode spacing,
and {\em iii)} the nucleotide configurations. Anyone of these changes 
can modify the values of the conductance, and even the relative conductance.

Zikic {\it et al.} also fail to mention that in our second work,
Ref.~\onlinecite{Lagerqvist2006-1}, we are sampling over more than a
thousand nucleotide configurations. Thus, one expects that sampling
over a non-random subset of configurations - as they do in their
work by only varying one angle - one can obtain different orderings
of the conductance. In fact, the alleged discrepancies instead
highlight and reinforce one of the main conclusions of our work (see
Ref.~\onlinecite{Lagerqvist2006-1}): in order to successfully
sequence DNA via transverse electronic transport, each device has to
be first calibrated by reading a known strand such that the current
distributions for all four nucleotides can be obtained. These
distributions are unique to {\em each and every} device and are
determined by the microscopic geometry of the pore and electrodes.

We now want to turn to two important questions: 1) why is it that
geometric factors are important in the conductance of nucleotides,
and 2) how can ``distinguishability survive averaging over possible
conformations of ss-DNA'' (using similar words to the question Zikic
{\it et al.} raise in their conclusions)?

The answer to the first question is something not stated by either
Zikic {\it et al.} or us. There are two factors that enable one to
focus mainly on the geometry of the nucleotide-electrode
configuration: 1) the HOMO and LUMO energies of the different bases
are close in energy compared to their distance from the gold Fermi
level. 2) For all four bases, the HOMO and LUMO states are
delocalized around the base, and thus one can substitute the atomic
structure of the bases as an approximate representation of the
spatial extension of the molecular states. Contrary to the
conclusions of Zikic {\it et al}., one can not say from their
results that HOMO and LUMO states are less important than geometry,
only that when comparing molecules of similar factors (1) and (2),
that the geometry would be the most dominant factor. Further, this leads
to a very important conclusion: in the case of sensors to detect the
DNA bases using electrical currents, the nucleotide configurations
have to be at least partially controlled. In terms of a nanopore-based device,
one way to do this is to use a transverse electric field induced by
the transverse electrodes or by an external capacitor across the
whole device.~\cite{Lagerqvist2006-1} Of course, there will be other
important factors to consider besides geometric fluctuations of the
nucleotides themselves, including the effects of ions.

\begin{figure}
\includegraphics*[width=7.5cm]{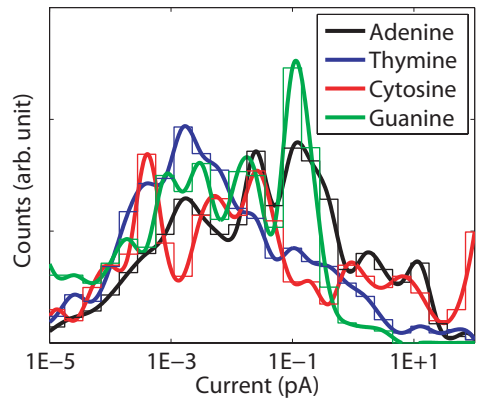}
\caption{Probability distributions of currents, for unstabilized
$\mathrm{poly(dX)}_\mathrm{15}$ as the strand propagates through a
pore with embedded electrodes. X is Adenine/Thymine/Cytosine/Guanine
for the black/blue/red/green curves, respectively. The thin lines
show the actual current intervals used for the count, while the
thick lines are an interpolation.} \label{distributions}
\end{figure}

The answer to the question on how the distinguishability survives by
averaging over possible conformations of ss-DNA, can be found by
examining, in the context of a nanopore-based device, the geometric
configurations of DNA as it translocates through a pore in the
absence of any control. This is the basis of our work in
Ref.~\onlinecite{Lagerqvist2006-1}. The geometrical fluctuations
cause the different nucleotides to have large fluctuations in the
value of their current, as shown in Figure~\ref{distributions}. With
no stabilizing transverse field added, the current distributions for
a given set of initial conditions of the different nucleotides
(calculated as reported in Ref.~\onlinecite{Lagerqvist2006-1}, for
an electrode bias of 0.1 V and an electrode spacing of 15
$\textrm{\AA}$) span several orders of magnitude, and have
significant overlap, as shown in the figure. These large
fluctuations will cause the bases to be essentially
indistinguishable. However, in the presence of a transverse field
that is much larger than the driving field, the nucleotides can be
stabilized, and thus distinguished by a relatively modest ensemble
measurement~\cite{Lagerqvist2006-1}.

We conclude by discussing the differences in using a tight-binding
(TB) approach compared to DFT calculations for the problem at hand.
Similar to the DFT approach within any available xc-functional, a TB
approach has its own limitations. Nonetheless, in the present
context it has a clear advantage. In particular, it satisfies two
conditions required by any method to investigate the relative
conductance of the nucleotides. First, since the coupling (i.e.,
geometry) is the large determining factor in the relative
conductance, one needs to adequately reproduce the spatial
distribution of the molecular wavefunctions. Second, the energies of
the molecular states need to be calculated fairly accurately. For
our chosen TB parameters, both of these quantities compare well with
DFT calculations for isolated nucleotides.
This, together with the
reduced computational complexity of TB, allows us to look at many
different geometric configurations to more realistically capture
structural fluctuations that would be present in an experiment.

In addition, there is no reason to believe that DFT, within the
xc-functionals used by Zikic {\it et al.}, can represent more
accurately the nucleotide-electrode coupling compared to TB in this
particular geometry where the nucleotides are not covalently bonded
to either electrodes.~\cite{Rydberg} Indeed, the
exchange-correlation functionals employed in Zikic {\it et al.}'s
work do not include the long-range van der Waals interactions that
are present in this weak coupling case. The fact that the two
different xc-functionals employed by Zikic {\it et al.} show order
of magnitude differences in the conductance may be a result of this
problem.

Finally, Zikic {\it et al.} state that ``As far as the
self-consistency of the electron transport is considered, this leads
to a procedure that is asymptotically exact in limit of zero
electric bias''. This is a misconception about static approaches to
transport, and is not correct even if one had the exact {\em static}
xc-functional. Two of the present authors (MZ and MD) have
shown~\cite{Sai} that even in the limit of zero bias, with the
inclusion of self-consistency, the current obtained using static DFT
{\em does not} include dynamical many-body effects which can only be
captured by using time-dependent approaches such as time-dependent
DFT.~\cite{Diventra} In Ref.~\onlinecite{Sai} we have evaluated
these dynamical corrections specifically for the local density
approximation functional, however, the statement is true regardless
of the static xc-functional chosen: no static xc-functional (even
the exact one) can fully capture the true non-equilibrium nature of
transport problems. Incidentally, these dynamical many-body effects
are also absent in the TB static approach to transport we have
employed.

\end{document}